\documentclass[conference]{IEEEtran}
\IEEEoverridecommandlockouts
\usepackage{times}
\usepackage{graphicx}
\usepackage{amsmath}
\usepackage{amssymb}
\usepackage{comment}
\usepackage{textcomp}
\usepackage[caption=false,font=footnotesize]{subfig}
\usepackage{multirow}
\usepackage{algorithm}
\usepackage{fixltx2e}
\usepackage[noend]{algpseudocode}
\usepackage{rotating}
\usepackage{booktabs}
\usepackage{lipsum}

\usepackage[outdir=./]{epstopdf}

\usepackage[pagebackref=false,breaklinks=true,letterpaper=true,colorlinks,bookmarks=false]{hyperref}

\usepackage{cite}

\ifCLASSINFOpdf
\else
\fi
\begin{document}
%
\vspace{-20pt}
\title{HyNNA: Improved Performance  for Neuromorphic Vision Sensor based Surveillance  \\using Hybrid Neural Network Architecture}

\vspace{-20pt}
\author{\IEEEauthorblockN{%
    Deepak Singla\IEEEauthorrefmark{1}\,
 Soham Chatterjee\IEEEauthorrefmark{1}\,
    Lavanya Ramapantulu\IEEEauthorrefmark{1}\,
    Andres Ussa\IEEEauthorrefmark{2}\,
    Bharath Ramesh\IEEEauthorrefmark{2}\,
    Arindam Basu\IEEEauthorrefmark{1}\thanks{Corresponding author: arindam.basu@ntu.edu.sg}
   }


 \IEEEauthorblockA{%
    \IEEEauthorrefmark{1}Nanyang Technological University 
    \IEEEauthorrefmark{2}National University of Singapore
 }} 
\vspace{-20pt}

%


\maketitle

\begin{abstract}
Applications in the Internet of Video Things (IoVT) domain have very tight constraints with respect to power and area. While neuromorphic vision sensors (NVS) may offer advantages over traditional imagers in this domain, the existing NVS systems either do not meet the power constraints or have not demonstrated end-to-end system performance. To address this, we improve on a recently proposed hybrid event-frame approach by using morphological image processing algorithms for region proposal and address the low-power requirement for object detection and classification by exploring various convolutional neural network (CNN) architectures. Specifically, we compare the results obtained from our object detection framework against the state-of-the-art low-power NVS surveillance system and show an improved accuracy of 82.16\% from 63.1\%. Moreover, we show that using multiple bits does not improve accuracy, and thus, system designers can save power and area by using only single bit event polarity information. In addition, we explore the CNN architecture space for object classification and show useful insights to trade-off accuracy for lower power using lesser memory and arithmetic operations.
\end{abstract}


%
\IEEEpeerreviewmaketitle

\vspace{-10pt}
\section{Introduction}

 
Internet of Things (IoT) applications are increasingly gaining traction leading to a demand for systems that can work with limited resources and in constrained environments~\cite{shi2016promise}. For example, on-field intelligent surveillance systems at road-sides junctures or government buildings need to use minimal area and power without a decrease in performance. This challenge is addressed by neuromorphic vision sensors (NVS) as they offer low-power sensing for Internet of Video Things (IoVT) applications. In contrast to a traditional RGB camera where all the pixels are sampled at fixed intervals, the pixels in an NVS fire asynchronously thus, offering higher transmission rate, temporal resolution and dynamic range in addition to low latency and power~\cite{DBLP:journals/corr/abs-1904-08405}. 

Typically, due to the real-time latency requirements of IoVT applications, object detection and recognition is performed at the sensor node itself. There are many systems using event-based cameras for object tracking\cite{drazen2011toward,delbruck2013robotic,glover2016event}, surveillance\cite{litzenberger2006estimation,pikatkowska2012spatiotemporal}, object recognition\cite{wiesmann2012event,DBLP:journals/corr/OrchardMEPTB15,lagorce2016hots,DBLP:journals/corr/abs-1803-07913}, gesture control\cite{amir2017low,lee2014real} and various other tasks. However, to the best of our knowledge, the system design challenge for an end-to-end NVS based surveillance system envisioned for IoVT application has not yet been addressed. 

While the recent study on a low-power end-to-end hybrid neuromorphic framework\cite{Authorsbmvc19} is the closest work that addresses system design for NVS based surveillance, it makes use of TrueNorth (TN) and FPGA hardware for recognizing moving objects, which consumes power in the order of watts while the power budget for IoVT applications is in the order of hundreds of milliwatts. Moreover, the incompatibility of TN deep learning models with other hardware platforms makes the widespread deployment of this system very challenging. In addition, this study uses a binary frame for object detection and classification, and thus, does not explore other forms of frame resolution. In contrast, our work uses the information from the NVS node similar to that of our brain's visual cortex diverging information into the ``what" and ``where" pathway, which has been previously tested on TN\cite{merolla2014million}. What pathway gathers the high resolution features of the image while the where pathway helps to locate salient objects.

\begin{figure}[!t]
\centering
\includegraphics[width=0.6\columnwidth]{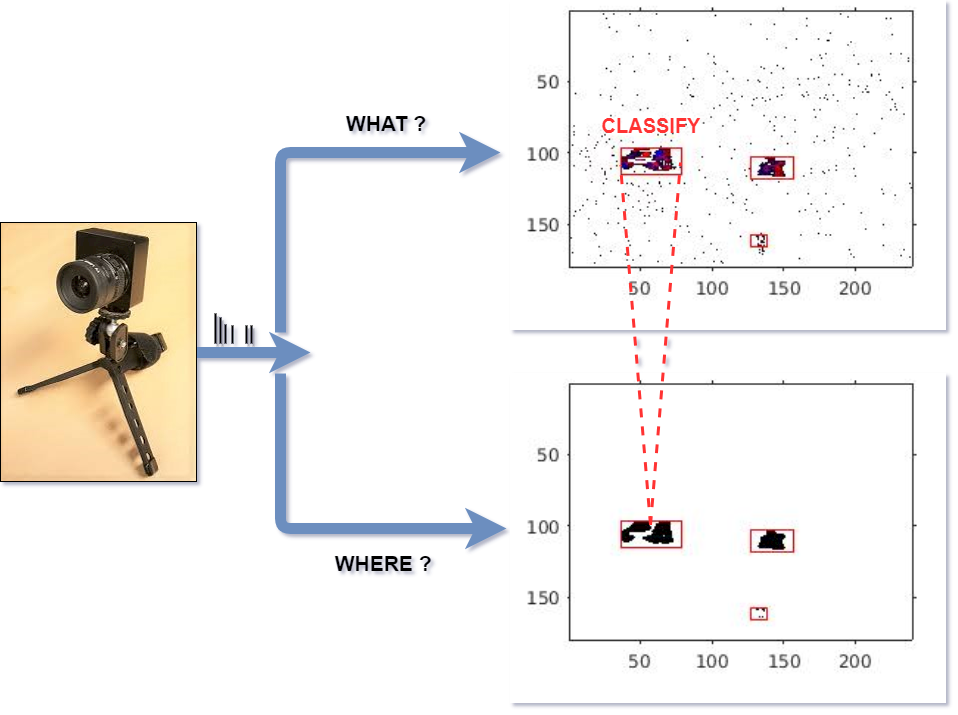}
\caption{Approach using what-where pathway with a DAVIS vision sensor} 
\label{fig_What_Where}
\vspace{-20pt}
\end{figure}

There are two types of paradigms that can be used to find salient information from events based on: individual events or groups of events. A single event does not provide much information for classification problems, hence most research works explore a hybrid approach where groups of events are accumulated into a frame either using a fixed number of events~\cite{DBLP:journals/corr/MoeysCKVDNKD16,lungu2017live} or a fixed interval~\cite{AuthorsEBBIOT}. These frames are passed through a deep learning framework during training. This trained network is converted back to a Spiking Neural Network (SNN) that processes data on an event-by-event basis\cite{DBLP:journals/corr/OrchardMEPTB15,lagorce2016hots,perez2013mapping,Authorsbmvc19}. Other works explore passing the same frames accumulated from temporal event information during both training and testing through classical deep learning frameworks in the form of dense, multi-channel tensors\cite{DBLP:journals/corr/MoeysCKVDNKD16,lungu2017live} which can provide better classification than the sub-optimal SNN like frameworks.
In contrast, this paper employs the approach of grouping events to a frame based on constant time and explore a complete scene with multiple moving objects. We use various levels of input feature complexity based on what-where pathway for classifying moving objects.  In addition, we explore different classical deep learning frameworks that are compatible for embedded hardware and can meet the demands of IoVT applications. Since we use the hybrid approach of event-frames~\cite{Authorsbmvc19}, we refer to our work as hybrid neural network architecture (HyNNA). HyNNA shows an improved performance of 19.06\% compared to the state-of-the-art\cite{Authorsbmvc19}. 
We elucidate the details of our approach in the following sections.
\vspace{-10pt}
\section{Methodology}
\vspace{-10pt}
\subsection{Overview}
The end-to-end software system design for NVS based surveillance uses three main modules: frame generation, object detection and object classification. Events from the NVS sensor node are converted into frames using the what and where pathway. Object detection is done on these generated frames using a region proposal algorithm. We discuss two different region proposal algorithms, histogram approach and the
connected component labeling approach. Lastly, we explore different convolutional neural network (CNN) architectures for object classification and observe useful insights on trade-off between accuracy and computations, and accuracy and memory requirements to save area and power.
\vspace{-5pt}
\subsection{Frame Generation}
We group NVS sensor events into frames by using two polarity channels based frames at fixed interval, $t_F$. We name the first channel: \textit{OFF-CHANNEL} and the second channel: \textit{ON-CHANNEL}, for storing the events when a log-intensity decrease or increase at the pixel is detected respectively. The events are stored as a tuple, $[e_i = (x_i, y_i, t_i, p_i)]$ where $(x_i,y_i)$ represent the event/pixel location on the sensor array, $t_i$ represent the timestamp of the event and $p_i$ represent the polarity ($p_i$ = 1 is ON event and $p_i$ = 0 is OFF event). These events are accumulated into image, $I_k$ where $k$ represents the frame number. Each event is given equal weightage and added to the previous spike record for a given frame $k$,
\vspace{-5pt}
\begin{align*}
\label{eq:accumulation}
    I_k(x_i,y_i,p_i) = I_k(x_i,y_i,p_i) + 1 
\end{align*}

where $p_i$ is 0 or 1. The spike record is clipped to a maximum of 15 spikes/pixel to limit spurious events from the NVS.
Such images are referred to as 
``multi-bit, 2 channel". If the spike record is clipped to one event per pixel per channel, such images are referred to as 
``1 bit, 2 channel". Lastly, both the channels are combined using logical-or operation making a 
``1 bit, 1 channel" image~\cite{Authorsbmvc19}. To utilize salient information, we use 
``multi-bit, 2 channel" for object classification (what) and 
{``1 bit, 1 channel"} for object detection (where), and map them using the what-where pathway as shown in Figure~\ref{fig_What_Where}.

\subsection{Region Proposal Algorithms for Object Detection}
Ussa et al.~\cite{Authorsbmvc19} implement a region proposal algorithm for object detection using two separate one-dimensional (1-D) histogram projections of the image along the X and Y axis, referred to in this paper as ``HIST RP". However, this approach leads to loss of signal information due to the mapping of the image to 1-D. 
Moreover, as this approach uses a sideview, regions for smaller objects get affected in the presence of bigger objects, making them the same size as the bigger objects along the given axis. To overcome these shortcomings, we propose to use a connected component labeling (CCL) approach for object detection. CCL is a two-dimensional (2-D) analysis of the image, and has been traditionally used as a two-pass algorithm for regular RGB image and video processing applications~\cite{he2017connected,walczyk2010comparative}.

As we are exploiting the visual cortex's what-where pathway approach, we use the ``1 bit, 1 channel" frame for object detection using CCL. In order to merge the nearby related pixels, we scale the frame size by downsizing it before applying CCL.
Downsizing also leads to a reduced number of computations, thus saving memory. Specifically, for the results in this paper, we apply a 6x3 logical-or patch to the ``1 bit, 1 channel" frame size of 240x180, thus reducing its size to 40x60. 
Using the regions of objects extracted from the CCL applied on a downsized frame, we map back these regions or bounding boxes (BBs) to the original image, thus producing BBs that are compact in size, and reducing the number of spurious features for the object classification module. We refer to this approach as ``CCL RP" when we discuss the results.


\subsection{CNN Architectures for Object Classification}
 \label{cnn_arch}
 Using what-where pathway approach, we use the ``multi-bit 2 channel" frame to derive the salient features necessary for object classification. While CNNs are  a popular choice for image classification, they cannot be used as is in the IoVT domain due to the lower power and memory constraints. To address this challenge, we explore different CNN architectures to reduce the number of computations and the memory footprint. We exploit the asynchronous events from NVS and the availability of a large frame time period, $t_F$. Secondly, we exploit the theory that most of the salient information in a classification network can be learnt within the first few layers, and thus, we can reduce the total number of layers in the CNN for an NVS based system. 
 
 Using lower number of layers, we start CNN exploration from a base LeNet5 architecture\cite{lecun1998gradient} consisting of three convolution (CONV) layers and two fully connected layers (FC). The convolution function is two-dimensional, but is performed for all the channels of the input image and the depth of the input volume in the following layers. To address the increasing number of computations by an order of magnitude with deeper depth dimensions, we decouple the convolution layers into depthwise separable convolutional (SEPCONV) layers using the Xception architecture~\cite{DBLP:journals/corr/Chollet16a}. In effect, it first involves independent channel-wise spatial convolution i.e. depthwise convolution, and then, projecting the channels output onto a new channel space using a pointwise convolution. While this has been explored for embedded vision applications in MobileNet architecture~\cite{DBLP:journals/corr/HowardZCKWWAA17}, it has not been explored for NVS applications with the usage of event-based frames over regular RGB frames. Using base Lenet5 architecture, we use the concept of depthwise followed by pointwise convolution (SEPCONV) for all the three convolution layers, and refer to this as a ``Base SepNet" architecture.
 
However, using only SEPCONV in all the first three layers (Base SepNet) loses accuracy as the feature maps are not used across all the channels. To overcome this loss of accuracy, we use a ``Mixed architecture" wherein regular CONV is used for the first layer, and then, SEPCONV for the next two layers along with FC layers as the last two layers of this network. While this may result in better accuracy, from the memory footprint perspective, all the three architectures, Base LeNet, Base SepNet and the Mixed architecture are the same. This is irrespective of whether the hardware is implemented using a layer-based approach where each network layer is processed sequentially or a tile-based approach where a tile of the input image is completely processed. This is counter-intuitive as tile-based implementations should result in a lesser memory footprint, but as the FC layer is the bottleneck for all the three network architectures, the memory footprint remains the same.

To overcome the memory bottleneck, we explore network architectures without any FC layer but, using only a softmax layer as the output layer to compute the confidence values of each class. We explored two different architectures with only CONV layers and no FC layer (Small LeNet), and a Mixed architecture with no FC layer (TinyNet) to exploit lower memory footprint and lower computations respectively. We show that in overall, the Small LeNet architecture is a good trade-off as reducing memory saves power and area without loss in accuracy compared to the TinyNet which has a lower accuracy. It is reasonable to trade-off the increased number of computations of Small LeNet with those of other CNN architectures because, the energy consumed by computations is an order of magnitude lower than that of memory accesses~\cite{horowitz20141}.

\section{Results}
In this section, we discuss the dataset extraction and experimental setup along with the evaluation results.
\subsection{Dataset Details}
The recordings from a DAVIS based NVS were taken on roads during daytime with two lens types: 12mm and 6mm. Frames were generated using $t_F = 66ms$, which is sufficient for traffic monitoring applications. The frame rate, frame noise filtering method and the dataset are same as the state-of-the-art work by Ussa et al.~\cite{Authorsbmvc19}.  These were manually annotated and a track ID was assigned to each and every different passing object in the frame. These Ground-Truth (GT) annotations were also assigned different classes such as car, bus, truck, van, human and bike. Human class was not considered in this study because it generates very less events, and does not provide enough features for classification, and it can be addressed with a longer $t_F$ value. 
The trucks and vans were combined due to their similar size and the trucks being lesser in number in the recorded scenes. Table \ref{tab:classdistribution} elaborates the dataset details for each class based on the GT annotations. As shown in the table, we have used objects of variety of sizes (in pixels) to illustrate the general application of our approach.
\vspace{-10pt}
\begin{table}[h]
\centering
\caption{class-wise distribution of data using GT boxes}
\label{tab:classdistribution}

\begin{tabular}{@{}llll|lll@{}}
\toprule
 & \multicolumn{3}{c|}{\textbf{Train/Validation}} & \multicolumn{3}{c}{\textbf{Test}} \\ \midrule
 & \textbf{\begin{tabular}[c]{@{}l@{}}Sample \\ Count\end{tabular}} & \textbf{\begin{tabular}[c]{@{}l@{}}Track \\ Count\end{tabular}} & \multicolumn{1}{c|}{\textbf{\begin{tabular}[c]{@{}c@{}}Size\\ WxH\end{tabular}}} & \textbf{\begin{tabular}[c]{@{}l@{}}Sample \\ Count\end{tabular}} & \textbf{\begin{tabular}[c]{@{}l@{}}Track\\ Count\end{tabular}} & \multicolumn{1}{c}{\textbf{\begin{tabular}[c]{@{}c@{}}Size\\ WxH\end{tabular}}} \\ \midrule
\multicolumn{1}{l|}{Car} & 17342 & 460 & 44x19 & 2984 & 84 & 44x19 \\
\multicolumn{1}{l|}{Bus} & 6954 & 246 & 101x41 & 1544 & 58 & 101x41 \\
\multicolumn{1}{l|}{Truck/Van} & 7272 & 205 & 50x25 & 1216 & 32 & 50x25 \\
\multicolumn{1}{l|}{Bike} & 1737 & 58 & 21x16 & 183 & 6 & 21x16 \\ \midrule
\multicolumn{1}{l|}{\textbf{Sum}} & 33305 & 969 &  & 5927 & 180 & 
\end{tabular}
\vspace{-10pt}
\end{table}

The outputs from the region proposal (RP) algorithms were matched with the GT annotations at the same time point using interpolation. We considered a RP output as a valid sample for classification only if it had an Intersection over Union (IoU as defined in \ref{eq:IoU}) greater than 0.1 with the GT box. 
\begin{equation*}
\label{eq:IoU}
    IoU=\frac{A_{Intersection}}{A_{Union}}
\end{equation*}
where $A_{Intersection}$ is the area of intersection and $A_{Union}$ is the area of union of RP box and the GT box.

As the CNN architectures that we explored require a fixed size of input tensor, we selected a 42x42 patch from the centroid of each box, from both the GT annotation or a valid RP output. The boxes which had size less than 42x42 were padded with zeros and adjusted to this common size. Further, in order to reduce the class-wise sample variance, and keeping in mind that the buses and trucks have larger mean size than 42x42, four patches from all the four corners of the box of buses and trucks were added to the train/validation dataset. This additionally helped the network to train on edges of the object since a RP might cover only half of the object. Thereafter, the bikes were rotated by random degrees within $\pm$15 and translated by some amounts to balance their numbers with the other classes. The data for testing also contained recordings from the same locations at different times and the same methodology was applied for getting samples from GT annotations as well as RP outputs.
\subsection{Experiment Details and Results}
\begin{table*}[h]
\caption{Classification accuracy using X/Y, where X is per-sample and Y is per-track overall balanced values}
\label{tab:my-table}
\resizebox{\textwidth}{!}{%
\begin{tabular}{|l|ccc|ccc|ccc|}
\hline
\textbf{} & \multicolumn{3}{l|}{\textbf{Train on GT, Test on GT}} & \multicolumn{3}{l|}{\textbf{Train on HIST RP, Test on HIST RP}} & \multicolumn{3}{l|}{\textbf{Train on CCL RP, Test on CCL RP}} \\ \cline{2-10} 
\textbf{CNN architectures} & \multicolumn{1}{c|}{\textbf{\begin{tabular}[c]{@{}c@{}}1 bit, \\ 1 channel\end{tabular}}} & \multicolumn{1}{c|}{\textbf{\begin{tabular}[c]{@{}c@{}}1 bit, \\ 2 channel\end{tabular}}} & \textbf{\begin{tabular}[c]{@{}c@{}}Multi bit, \\ 2 channel\end{tabular}} & \multicolumn{1}{c|}{\textbf{\begin{tabular}[c]{@{}c@{}}1 bit, \\ 1 channel\end{tabular}}} & \multicolumn{1}{c|}{\textbf{\begin{tabular}[c]{@{}c@{}}1 bit, \\ 2 channel\end{tabular}}} & \textbf{\begin{tabular}[c]{@{}c@{}}Multi bit, \\ 2 channel\end{tabular}} & \multicolumn{1}{c|}{\textbf{\begin{tabular}[c]{@{}c@{}}1 bit, \\ 1 channel\end{tabular}}} & \multicolumn{1}{c|}{\textbf{\begin{tabular}[c]{@{}c@{}}1 bit, \\ 2 channel\end{tabular}}} & \textbf{\begin{tabular}[c]{@{}c@{}}Multi bit, \\ 2 channel\end{tabular}} \\ \hline
\textbf{Base Lenet5} & 86.47/93.48 & \textbf{89.64/94.77} & 87.6/92.91 & 71.15/81.80 & 75.26/79.94 & 74.34/81.02 & 78.36/91.21 & \textbf{82.16/93.56} & 81.92/89.65 \\
\textbf{Base SepNet} & 85.87/92.48 & 87.33/92.72 & 87.31/92.43 & 72.06/81.80 & \textbf{74.76/86.15} & 74.72/81.32 & 77.41/91.51 & 81.18/91.21 & 81.19/90.43 \\
\textbf{Mixed Architecture} & 85.53/92.83 & 86.98/92.91 & 87.59/90.08 & 71.58/87.23 & 74.54/82.29 & 72.87/80.74 & 76.50/92.48 & 81.22/93.56 & 81.61/92.78 \\
\textbf{Small LeNet} & 85.39/92.35 & 87.55/91.94 & 85.88/93.56 & 69.11/80.61 & 72.56/79.64 & 71.52/76.89 & 76.05/87.53 & 81.64/94.34 & 80.02/89.65 \\ \hline
\end{tabular}

}
\vspace{-10pt}
\end{table*}

Experiments were done with images using three resolutions: ``{1 bit, 1 channel}", ``{1 bit, 2 channel}" and ``{multi-bit, 2 channel}". Mini batch stochastic gradient descent with Adam optimiser~\cite{kingma2014adam} was used for training. Data was shuffled and fed into the network in batches of size 128. Models were trained in a Keras framework with NVIDIA TITANX GPU for 20 epochs and the model with the highest validation accuracy was saved. To tackle the class samples imbalance, we evaluated our results in the form of overall balanced accuracy: per-sample is the average of class-wise sample accuracies and per-track is the average of class-wise track accuracies. The label for a track is correctly predicted, only if the mode predicted label for all the samples in the track is same as the actual label for the track.\footnote{The voting is based on the ground-truth tracker. This is similar to the state-of-the-art with which we compare our results.} While the Base Lenet5 architecture achieved the highest accuracy of 82.16\%, it also required the highest number of FLOPS and memory. Further, single bit images with polarity information perform better than multi-bit images, thus helping the designers to choose low power/area. Table \ref{tab:my-table} shows the results of the CNN architectures that resulted in either reduction in computations or memory but also reducing accuracy for the architectural choices described in Section~\ref{cnn_arch}.
\subsection{Memory and FLOPS of CNN architectures}
\begin{figure}[htbp]

\vspace{-15pt}
\begin{minipage}[l]{1.0\columnwidth}
\subfloat[FLOPS vs. Accuracy]{%
 \includegraphics[width=0.49\columnwidth]{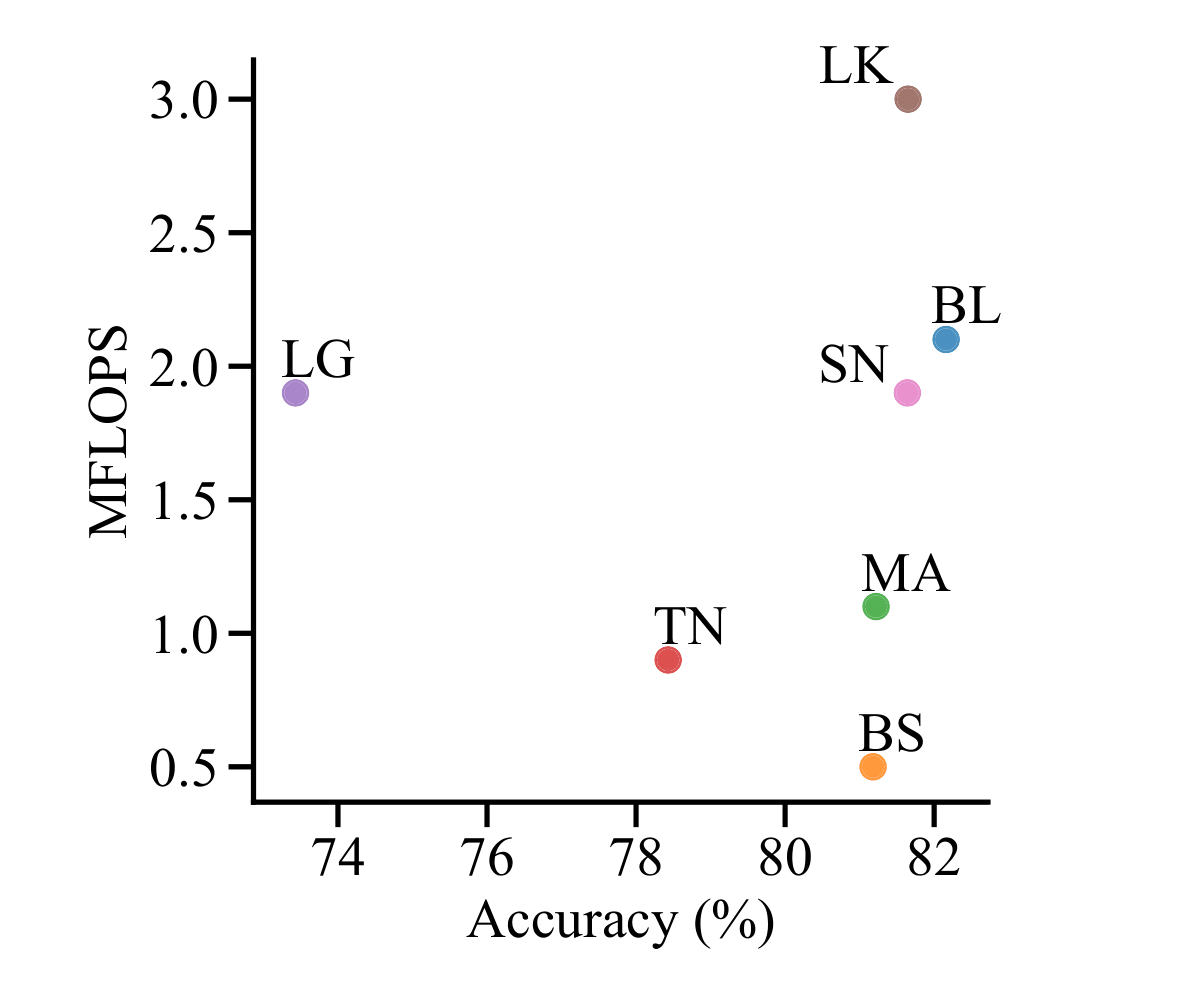}
}
\subfloat[Memory vs. Accuracy]{%
  \includegraphics[width=0.49\columnwidth]{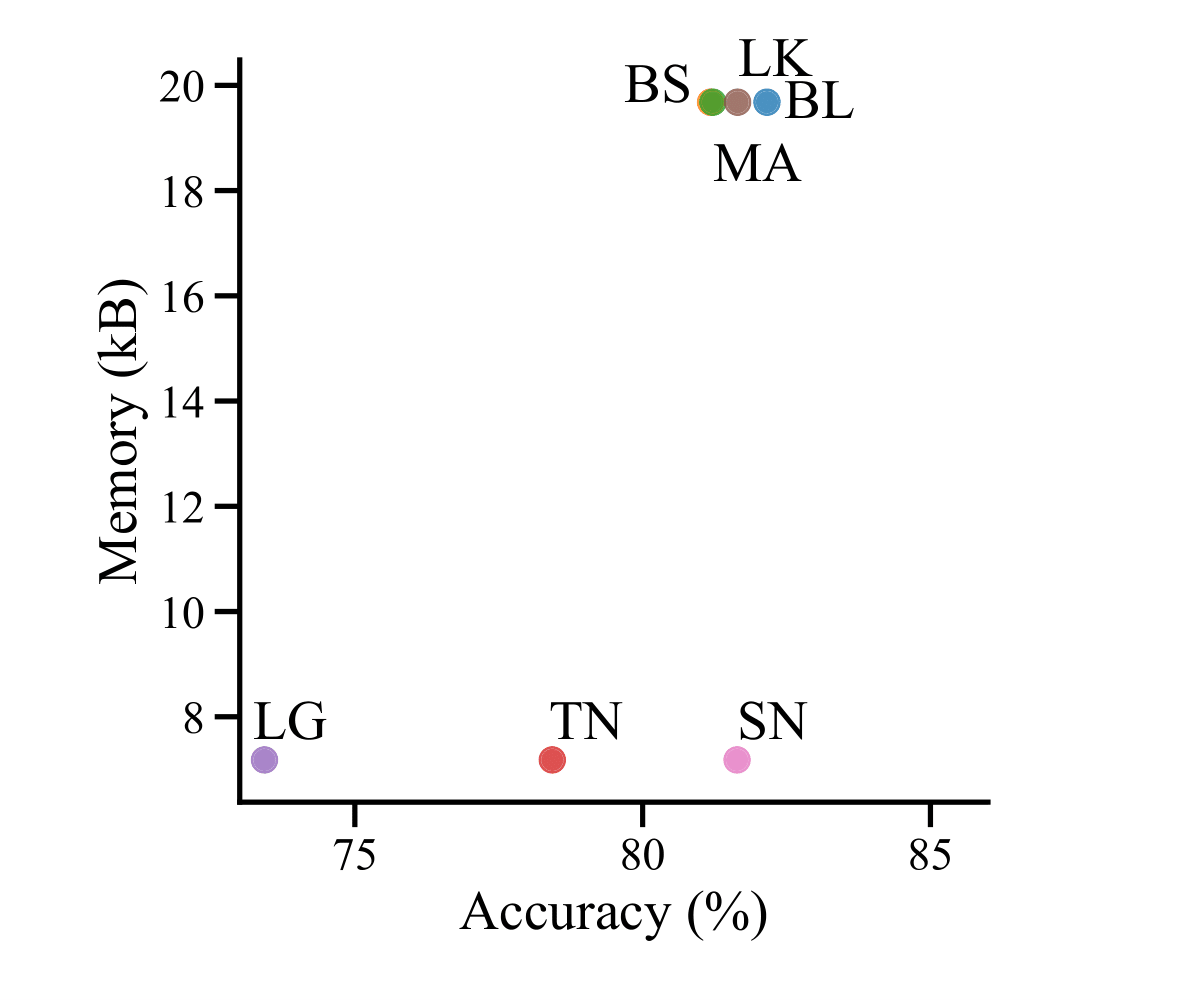}%
}
\caption{FLOPS, Memory and Accuracy Trade-offs for CNN Models}
\label{fig:memory accuracy}
\end{minipage}
\vspace{-10pt}
\end{figure}

To choose an appropriate CNN architecture for an NVS system design is non-trivial due to the innumerable design choices available, and the non-obvious impacts of these choices on accuracy, hardware area and power. To aid NVS system designers garner insights on trade-off choices, Figure~\ref{fig:memory accuracy} shows results for accuracy versus number of computations and memory required for the CNN architectures trained using the ``1 bit, 2 channel CCL RP" data. The memory size is  calculated using a tile size of 21 being propagated across all the layers instead of processing each layer sequentially.\footnote{Decreasing the tile size reduces memory but increases latency.} Analyzing the CNN architectures towards the bottom right corner of the figures, we observe that the Small LeNet architecture is a good choice as it reduces the memory usage by 63.51\% with a drop in accuracy by 0.52\%. The savings in memory reduce energy by an order of magnitude compared to architectures that reduce computations as shown by Horowitz et al~\cite{horowitz20141}.




\begin{table}[]
\vspace{-10pt}
\caption{Description of CNN architectures}

\resizebox{0.99\columnwidth}{!}{
\begin{tabular}{lll}
\hline
\multicolumn{1}{l}{\textbf{Label}} & \multicolumn{1}{l}{\textbf{Architecture}} & \multicolumn{1}{l}{\textbf{Hyperparameters}}                                                                                                                         \\ \hline
BL                                 & Base LeNet5                                & \begin{tabular}[c]{@{}l@{}}Base LeNet5 architecture with two\\ CONV layers with average pooling \\after each layer and three FC layers.\end{tabular} \\ \hline
BN                                 & Base SepNet                               & \begin{tabular}[c]{@{}l@{}}LeNet Architecture  with SEPCONV \\ instead of 2D CONV layers \end{tabular}                                               \\ \hline
MA                                 & Mixed Architecture                        & \begin{tabular}[c]{@{}l@{}}LeNet Architecture with  second  layer \\ as SEPCONV instead of 2D CONV \end{tabular}                                     \\ \hline
TN                                 & TinyNet                                   & \begin{tabular}[c]{@{}l@{}} Mixed architecture\\ with 5 filters in second layer\\ and only one softmax dense layer\end{tabular}                                 \\ \hline
LG                                 & LeNet With Global Pooling                 & \begin{tabular}[c]{@{}l@{}}Base LeNet with Global Average\\ Pooling instead of Flatten layer\end{tabular}                                                            \\ \hline
LK                                 & LeNet Large Kernel                        & \begin{tabular}[c]{@{}l@{}}Base LeNet with 7x7 kernels in\\ all 2D CONV layers\end{tabular}                                                                   \\ \hline
SN                                 & Small LeNet                               & \begin{tabular}[c]{@{}l@{}}Base LeNet with only one\\ Softmax dense layer\end{tabular}                                                                               \\ \hline
\end{tabular}
}
\vspace{-10pt}
\end{table}

\subsection{Comparison to state-of-the-art}
To show a fair comparison with state-of-the-art by Ussa et al.\cite{Authorsbmvc19}, we compared classification accuracy from our CNN models to per-sample TN test accuracy with samples from FPGA tracker as the train data. As we did not use human and other (background samples) class in our experiment, we recalculated the overall balanced accuracy from the paper without these two classes. We found that there is a major improvement from 63.13\% to 82.16\% using Base Lenet as the CNN model with ``{1 bit, 2 channel}" frame as the input. Even for the ground truth train/test data, the per track accuracy improves by 6\% from 87.75\% to 94.77\%. This shows that using an additional channel with polarity and using a simple CNN model leads to a better accuracy compared to using a ``{1 bit, 1 channel}" and TN based deep learning models for NVS based surveillance.

\vspace{-10pt}
\section{Conclusion}
This paper proposes a hybrid neural network architecture based approach, HyNNA for NVS based surveillance applications. HyNNA generates frames by considering both the on and off polarity events as two image channels. Comparing HyNNA against the state-of-the-art, we show an improved accuracy of 82.16\% due to usage of dual channel images and CNN architecture for classification. Moreover, we show that using multiple bits does not improve accuracy, and thus, system designers can save power and area by using only single bit polarity information of events. In addition, we explore different CNN architectures to trade-off accuracy and power/area, and show that using a mixed architecture CNN can achieve savings of 44.87\% in computations with no loss of accuracy. Furthermore, by removing the last few fully-connected layers in the network and using a tiling architecture, memory savings up to 63.51\% can be achieved with only a 0.52\% loss in classification accuracy.


\section*{Acknowledgment}
We would like to thank the authors of ~\cite{Authorsbmvc19} for sharing their dataset and results to compare our approach with their work.

\bibliographystyle{IEEEtran}

\bibliography{main_file}

\end{document}